\documentclass{IEEEcsmag}

\usepackage[colorlinks,urlcolor=blue,linkcolor=blue,citecolor=blue]{hyperref}
\expandafter\def\expandafter\UrlBreaks\expandafter{\UrlBreaks\do\/\do\*\do\-\do\~\do\'\do\"\do\-}
\usepackage{upmath,color}
\usepackage{cite}
\usepackage{url}

\pubyear{2025}

\newif\ifdraft
\drafttrue
\ifdraft
  \newcommand{\todo}[1]{{\textcolor{red}{ TODO: #1 }}}
  \newcommand{\ian}[1]{{\textcolor{green}{ Ian: #1 }}}
  \newcommand{\greg}[1]{{\textcolor{green}{ Greg: #1 }}}
  \newcommand{\kyle}[1]{{\textcolor{teal}{ Kyle: #1 }}}
\else
  \newcommand{\todo}[1]{}
  \newcommand{\ian}[1]{}
  \newcommand{\greg}[1]{}
  \newcommand{\kyle}[1]{}
\fi

\begin{document}

\title{Agentic Discovery: Closing the Loop with Cooperative Agents}

\author{J.~Gregory~Pauloski}
\affil{University of Chicago, Chicago, IL 60637, USA}

\author{Kyle~Chard}
\affil{University of Chicago, Chicago, IL, 60637 and Argonne National Laboratory, Lemont, IL 60439, USA}

\author{Ian~T.~Foster}
\affil{Argonne National Laboratory, Lemont, IL 60439 and University of Chicago, Chicago, IL 60637, USA}

\begin{abstract}
As data-driven methods, artificial intelligence (AI), and automated workflows accelerate scientific tasks, we see the rate of discovery increasingly limited by human decision-making tasks such as setting objectives, generating hypotheses, and designing experiments.
We postulate that cooperative agents are needed to augment the role of humans and enable autonomous discovery.
Realizing such agents will require progress in both AI and infrastructure.
\end{abstract}

\maketitle

\chapteri{M}odern research demands the seamless integration of experiments, observations, models, simulations, artificial intelligence (AI), and machine learning (ML)---all while processing ever-growing volumes of data and leveraging large, diverse, and distributed computing infrastructure.
This situation is emblematic of broader transformations associated with the fourth and fifth paradigms of science, which capture the shift towards data-intensive methods and artificial intelligence, respectively, as integral aspects of scientific exploration~\cite{hey2009fourth,malitsky2018fifthp}.
Fields ranging from astrophysics to social sciences now rely on vast datasets, AI models, and computational methods to drive innovation.
Hence we face the challenge of not just managing data and building models, but also building systems that enable researchers to integrate and utilize data and models at scale.
Current approaches to integrating data-intensive workflows and AI methods have yielded successes, but use techniques that result in siloed solutions that fail to scale or generalize.
This paradigm shift demands more than building increasingly sophisticated tools; it calls for a fundamental rethinking of how science is conducted.

Many aspects of today's iterative research process can be automated (data acquisition and preparation, workflow orchestration, modeling and simulation, and data visualization); however, human experts are still required to propose hypotheses, design experiments, write programs, procure resources, and interpret results.
These human-driven aspects require experts to keep up with state-of-the-art concepts, techniques, and results---a task that is increasingly impractical due to exponential growth in published data and papers~\cite{hanson2024publishing}.
Sustaining the ever-growing scale of scientific endeavors will require addressing these human limitations.

Agents---programs capable of independently or semi-autonomously performing tasks on behalf of a client or another agent---have been studied for decades, but contemporaneous advancements in AI have reinvigorated interest.
Namely, the reasoning capabilities of large language models (LLMs) have improved to the point where intelligent agents can steer more complex processes with the flexibility and autonomy that previously only humans could provide.
A future in which agents play more anthropomorphic roles in iterative research cycles is tangible.
Like human experts, agents are typically specialized and share the autonomous, persistent, stateful, and collaborative attributes necessary to work in unison to achieve broader goals.
Yet, specialized agents can perform their roles more efficiently than humans---at least to an extent---enabling more autonomous, higher throughput, and arguably more reliable discovery.

As a programming paradigm, agents can serve as both a valuable conceptual model and a practical framework for integrating loosely coupled systems that compose large-scale research infrastructure. As such, they can be the missing piece that enables long-running, autonomous use of these resources.
An agent-driven future of science requires continued advancements at many levels of abstraction and attention to both human and technical challenges, but we posit that the payoff will be immeasurable. 

In this article, we review the history
of agents from conceptualization with actors through to modern adoption in AI-based ``agentic workflows.'' We describe our prediction for an era of agentic discovery in which federations of agents work cooperatively to augment or replace humans within scientific processes. We present a case study in materials discovery to orient our prediction and describe how agents will transform every phase of scientific discovery to create an autonomous discovery process. We distill key technical challenges that must be addressed to achieve such visions and outline several uncertainties and risks.

\section{What is an Agent?}

Agent-based frameworks have proliferated recently; yet, the reemergence of agents in LLM contexts obscures the breadth of research encompassed by the term.
Contextualizing the history, taxonomy, successes, and failures of agents is key to evaluating the impact of agentic applications for scientific discovery.

The origins of agentic systems are rooted in Carl Hewitt's actor model~\cite{hewitt1973actors}.
Actors are independent computational entities that enable concurrent computing (where the lifetimes of many distinct computations overlap) through asynchronous message passing.
In response to a message, an actor can alter its local state, send messages to other actors, and create new actors.
This conceptual model is simple---lack of global state obviates the need for locks and other synchronization primitives---and powerful---Hewitt describes that ``all of the modes of behavior can be defined in terms of one kind of behavior: sending messages to actors.''
Laying a formal foundation for distributed systems, the actor model influenced many early concurrent programming languages (e.g., inter-process communication via mailboxes in Erlang and tuple-spaces in Linda) and methods for fault-tolerance, state management, and consistency (e.g., Paxos~\cite{lamport1998paxos}).

A formal definition of ``agent'' has been long sought.
Abstractly, an agent is an entity---something or someone---that acts on behalf of another entity; this is rather actor-like within the context of computer science.
However, popular use of the term ``agent'' emerged throughout the 1980s within distributed artificial intelligence.
Then, and continuing into the 1990s, AI was often regarded as ``the subfield of computer science which aims to construct agents that exhibit aspects of intelligent behaviour'' \cite{wooldridge1995agents}.
For example, reinforcement learning methods such as Q-learning~\cite{watkins1992q} considered how to develop agents that maximize a reward through interacting with the environment.
Like actors, agents were entities that could communicate and manage their own internal state, but exhibited intelligent behavior weakly defined as operating autonomously, perceiving and reacting to the environment, and taking goal-oriented actions.

While modern AI methods are dramatically different, multi-agent systems (MAS) were seen as critical to distributed artificial intelligence.
Researchers posited that a MAS would exhibit emergent behavior~\cite{walker1995emergence}, enabling more complex problem solving through internal decision making where agents collaborate by goal setting, planning, negotiating, and reasoning, as in BDI (belief, desire, and intention)~\cite{rao1997bdi}.

Despite the theoretical promise of MAS, practical challenges arose.
Scalability, robustness, and the complexity of designing intelligent, cooperative agents hindered widespread adoption in the early 2000s. 
While some successes were notable, such as agent-based modeling, MAS struggled to deliver on the ambitious vision of emergence at scale.
The actor model remained relevant within distributed computing, but the field of AI was no longer defined by the practice of constructing agents following the advancements in hardware and algorithms that lead to the rise of deep learning.

Agents found renewed relevance in the 2020s with the advent of LLMs. 
These systems, pre-trained on vast corpora of text, exhibit emergent capabilities in natural language understanding and reasoning. 
By embedding LLMs within agent architectures, researchers revisited earlier visions of autonomous, intelligent agents with newfound capabilities. 
Frameworks like AutoGen~\cite{wu2023autogen} and OpenAI Swarm~\cite{openai-swarm} leverage LLMs to create agents that can perform tasks such as information retrieval, summarization, and interactive collaboration with humans and other agents.
Tool calling, such as in Claude~\cite{claude2025function} or LangChain~\cite{langchain2025function}, enables LLMs to invoke user-provided external functions or APIs.
The AI Scientist~\cite{lu2024aiscientist} is an autonomous agent that carries out the entire research process, from hypothesis generation to paper writing, in an iterative, tool-assisted loop.
Google's AI co-scientist~\cite{google2025coscientist}, built on Gemini 2.0, is a multi-agent AI system designed to collaborate with scientists by generating novel hypotheses and research proposals.
In these classes of MAS, each agent plays a special role defined by its system prompt; agents then communicate, iteratively refining the response or mapping messages to actions, to satisfy their goal---the client's query.

Within computer science, an agent remains---and will likely remain---ill-defined.
Rather than attempt to naively reconcile this history through an $(n+1)$th definition of an agent (c.f.\ \href{https://xkcd.com/927/}{xkcd 927} for a humorous take), we contextualize the remaining discussion by summarizing the kinds of high-level behaviors that an agent can exhibit.
An agent is broadly classified as \emph{deliberative} (often called \emph{intelligent}) or \emph{reactive}.
A deliberative agent contains a model of the environment (a state) that is used to reason about what long-term plans to make in order to achieve its goal~\cite{wooldridge1995agents}.
In contrast, reactive agents lack a world model and only take actions in response to changes in their perceived environment~\cite{nwana1996agents}.

We can further define more specialized behaviors.
\emph{Service} agents provide predefined services, including executing computational routines or providing resource access.
\emph{Embodied} agents interact with the physical world, often via an actuator.
\emph{Learning} agents refine their behavior over time, typically leveraging reinforcement learning to enhance performance.
\emph{AI} agents employ an AI model, such as an LLM, for taking actions or making decisions.
A MAS is composed of \emph{cooperative} agents that work together toward high-level system goals by planning and coordinating smaller tasks across agents with the appropriate behaviors and resources.
To a client, a MAS can manifest as a single agent, encapsulating the complexity of the system.
We emphasize that nothing precludes an agent from exhibiting multiple behaviors---in fact, most do!

\section{An Era of Agentic Discovery}

We predict that federations of cooperative agents---deliberative, reactive, embodied, and more---will augment, and often replace, humans in the loop in scientific endeavors.
This prediction is motivated by two observations.
First, human decision making often limits the rate of discovery; second, advances in agentic systems are converging to a point where the complete scientific method can be carried out autonomously.
We explore the first observation through a case study in the autonomous discovery of carbon capture materials.
Then we describe a future in which specialized agents champion each phase of the scientific method.

\subsection{A Case Study in Materials Discovery}

\begin{figure*}
    \centering
    \includegraphics[width=1\linewidth]{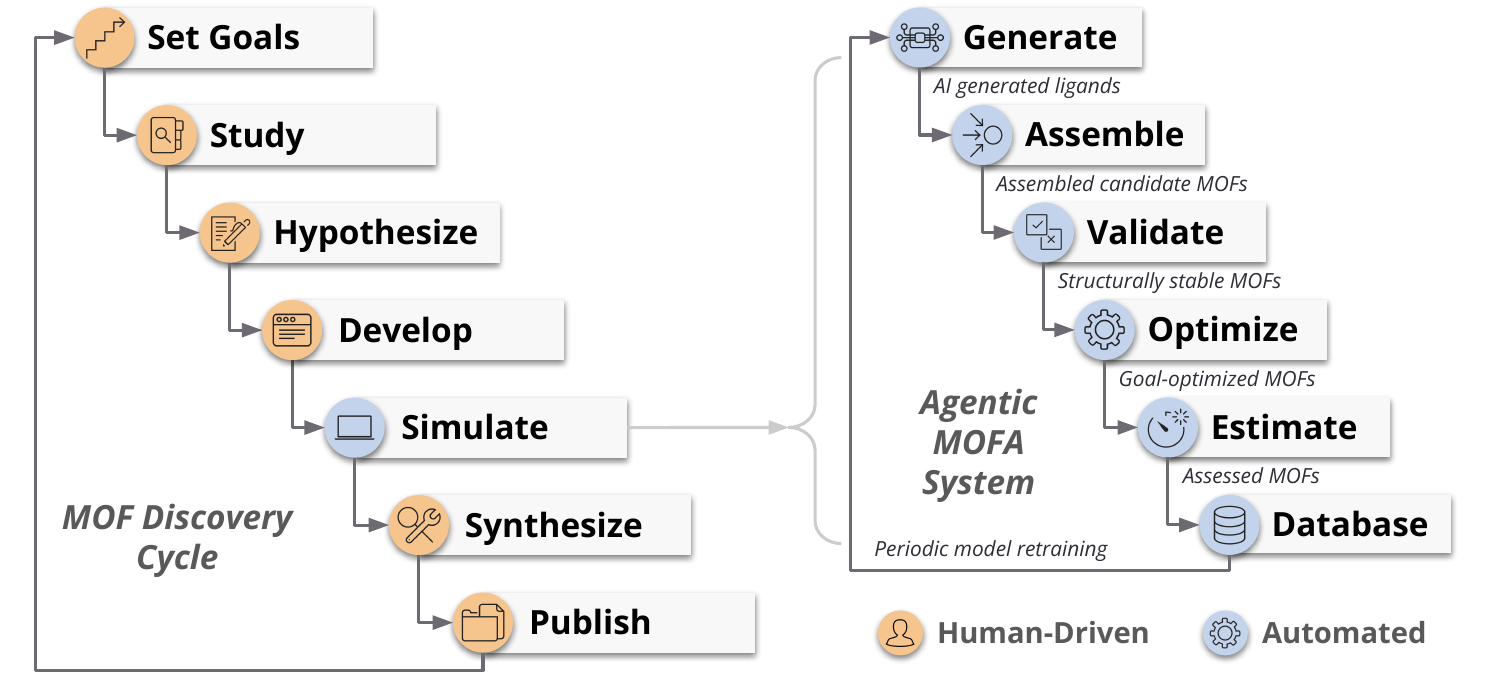}
    \caption{
    The discovery cycle of metal-organic frameworks (MOFs) for carbon capture is largely human-driven (orange stages). While some aspects have been automated (blue stages), such the AI generation and simulation of MOFs in the agentic MOFA system, human responsibilities limit the rate of MOF discovery.
    }
    \label{fig:mofa}
\end{figure*}

The urgency of climate change demands innovative solutions across multiple fronts.
Carbon capture is a crucial component of this multi-targeted approach, as reducing atmospheric CO$_2$ levels can mitigate the greenhouse effect and slow global warming.
Traditional methods of carbon capture, such as chemical scrubbing and geological storage, face challenges in efficiency, scalability, and cost~\cite{ma2022carbon}.
Metal-organic frameworks (MOFs) offer a promising alternative; these polymers, composed of inorganic metal clusters and organic ligands, are porous with high surface area.
MOFs can be tuned for selective gas adsorption properties making them highly effective for capturing and storing CO$_2$.
Discovering new MOFs with optimal properties is a daunting task due to the intractable combinatorial space of possible structures and the costs associated with synthesis and evaluation.
Thus, scientists desire autonomous methods for screening vast quantities of MOF structures.
One such example is MOFA~\cite{yan25mofa}, an online learning framework for generating, screening, and evaluating MOFs that couples generative AI methods with computational chemistry.

The MOFA workflow is as follows: (1) a specialized AI model generates candidate ligands, (2) candidate MOFs are assembled using predefined metal clusters and generated ligands, (3) candidates are iteratively screened and validated using multiple molecular dynamics computations, (4) CO$_2$ adsorption of promising candidates is simulated and stored in a database, (5) the generative AI model is periodically retrained on these results to improve performance.
MOFA is representative of a broad class of scientific workflows that have contributed to advancements in many fields.
While these workflows aim to enable autonomous discovery, their rigid and tightly coupled design means that humans are still primarily responsible for stewarding their utility in broader scientific endeavors.

MOFA can screen MOFs many orders of magnitude faster than any human could synthesize and evaluate them, having been demonstrated to identify thousands of stable MOFs per hour.
Yet, MOFA represents only a small step within the search for better carbon capture materials so the benefits of this acceleration are poorly realized.
Consider the steps that happen before or after a scientist executes MOFA (depicted in \autoref{fig:mofa}): domain experts distill techniques and results from the literature to inform hypotheses, programmers design more accurate and faster molecular dynamics simulations, ML practitioners investigate new generative model architectures, and chemists synthesize and evaluate MOFs in laboratories.
The outcomes of each step influence each other, but propagation and application of outcomes is human-driven which introduces considerable latency.
Accelerating an individual task is good, such as using MOFA for high-throughput screening, but accelerating decisions can have far greater impacts.
Multi-agent systems are ideal for optimizing these research processes, retaining the autonomy of different research components while automating decision making.

An agentic version of the MOFA workflow, built on the Academy federated agents framework, demonstrates early benefits of such agentic workflows~\cite{pauloski2025academy}.
The agent-based architecture naturally accommodates asynchronous execution and decentralized decision-making, making the workflow more resilient to variable workloads, resource fluctuations, and agent availability when deployed across federated environments.
While this initial approach shows clear promise, extending the agentic workflow to fully encompass all phases of the carbon capture materials research cycle offers the potential for even greater acceleration and impact.

\subsection{Closing the Loop with Agents}

\begin{figure*}
    \centering
    \includegraphics[width=1\linewidth]{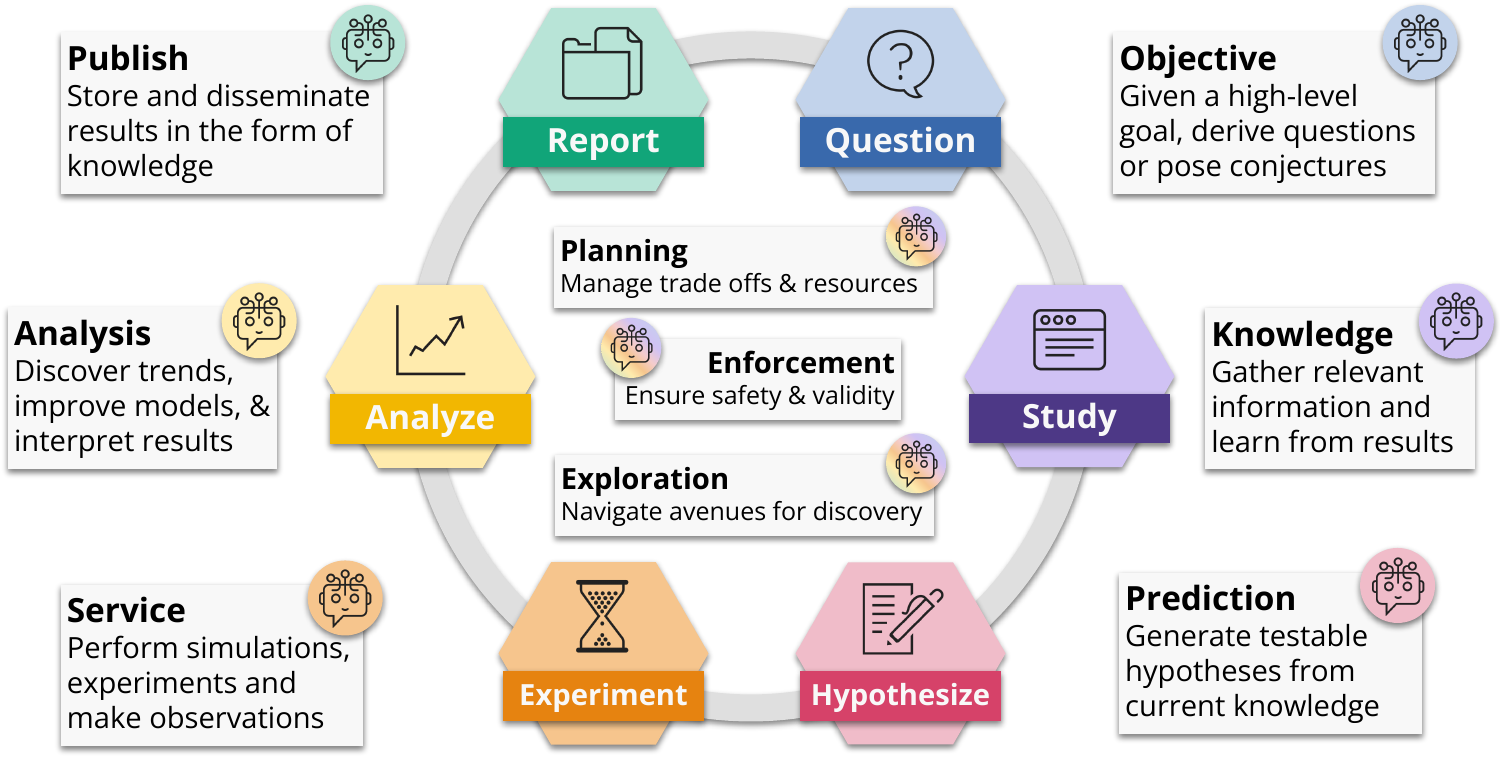}
    \caption{
    The scientific method is an iterative process (stages depicted in the central loop).
    Specialized agents (depicted as boxes with corresponding stages indicated by color) can carry out the stages autonomously.
    Agents can also transcend stages to enable long-term planning, exploration, and safety.
    }
    \label{fig:discovery}
\end{figure*}

Scientific discovery is inherently iterative, involving goal definition, research, hypothesis generation, experimentation, analysis, and dissemination.
Agents can play specialized roles at each phase while also transcending individual steps to coordinate broader research objectives.
We discuss each of these roles in detail, including examples within our materials discovery use case, and summarize this ecosystem of cooperative agents in \autoref{fig:discovery}.

We show in the figure a set of intelligent learning agents that transcend any one phase of the method and instead guide the actions of other agents in the MAS. 
An \textbf{Exploration} agent steers the system, prioritizing breadth in early phases then transitioning towards more targeted exploration when promising avenues are discovered. 
A \textbf{Planning} agent is responsible for managing the environment and inherent tradeoffs in which the agent operates, for example, to guide allocation of shared resources, follow predefined policies, and so forth.
An \textbf{Enforcement} agent plays the critical role of ensuring that agents' actions are safe, legal, and meet other regulatory requirements. 
These agents interact with many, if not all, of the agents in each phase.
\textit{In MOFA, agents can steer exploration of vast chemical design spaces, plan long-running experiments, allocate resources across specialized agents, and ensure that actions satisfy safety and technical constraints.}

The scientific process is framed around a particular goal. For example, to discover MOFs that are effective at capturing and storing CO$_2$, to understand the nature of the universe, or to characterize the molecular pathways that lead to a specific disease. As such, the first phase of the method aims to define, with some degree of specificity, the question(s) that guide the subsequent phases. An \textbf{Objective} agent, when given a high level goal by a human, can derive a set of questions or pose conjectures.
Enabling agent-to-agent or agent-to-human debate can improve the deliberation process of these agents, producing better or more refined questions.
\textit{In MOFA, an LLM-based AI agent can conjecture about possible base metal nodes of interest or ask questions about the performance impacts of specific structure geometries.}

Given a set of questions, the next phase in the process, study, seeks to identify knowledge that can help address the question.
The \textbf{Knowledge} agent must mine literature, identify relevant prior experimental and simulation results, obtain published data, and establish linkages between potentially related information.
\textit{In MOFA, a knowledge agent can utilize retrieval retrieval-augmented generation (RAG) to investigate prior uses of metal nodes of interest, or leverage embedding databases to find similar structures of interest.}

The \textbf{Prediction} agent
synthesizes the questions, conjectures, and knowledge from prior agents into hypotheses that can be tested.
This agent learns over time and incorporates a degree of creativity in its decision making.
A key aspect of the learning process is improving the feasibility of hypotheses using feedback from experimental agents.
\textit{In MOFA, the prediction agent proposes changes to certain input parameters that will result in a desired effect.}

The experiment phase seeks to gather data that can prove or disprove hypotheses.
Experiments, conducted by one or more \textbf{Service} agents, may encompass both physical experiments conducted via automated laboratories (embodied agents), simulated experiments on HPC infrastructure (computational or code-generation agents), and observational measurements from sensors (observational agents).
Service agents may self-coordinate in a peer-to-peer fashion, or engage planning agents to break up tasks into smaller actions to be dispatched to appropriate agents.
\textit{In MOFA, AI agents can generate candidate ligands, computational agents can perform screening simulations, and embodied agents can synthesize materials and evaluate them within self-driving laboratories.}

Data and observations from experiments are interpreted by \textbf{Analysis} agents.
These agents look for trends or patterns in the data, may train or use models, and will ultimately interpret the data to derive findings.
\textit{In MOFA, statistical analysis and causal inference agents can use results to determine the veracity of hypotheses and review the efficiency of performed experiments (e.g., which assays were most indicative, resource utilization, etc.).}

Finally, the \textbf{Publish} agent will store and disseminate the results in the form of knowledge. Depending on the consumer, this knowledge may be communicated in different ways, for example writing to an embedding database or knowledge base for other agents, preparing a video to share with the public, or publishing a paper to share with experts.
Importantly, this agent must capture the provenance of the research processes in a way that the results are verifiable, understandable, and ideally, reproducible.
\textit{In MOFA, the publish agent can write generated MOFs to a database and disseminates outcomes to the objective, knowledge, and prediction agents.}

\subsection{Evolving Human Responsibilities}

Realizing this world of agent-driven discovery does not negate the need for nor the utility of scientists.
Rather, we envision that the responsibilities of scientists will transition away from mundane tasks, such as experiment execution, resource provisioning, and monitoring, to higher-level objectives.
Strategic decision making, long-term objective setting, theoretical and conceptual development, interdisciplinary collaboration, verification and validation of findings, and general system design remain central tasks for scientists.
The key distinction with these tasks is that they are not responsible for the overheads that impact the rate of discovery.

\section{Key Technical Challenges}

Achieving this vision of fully autonomous discovery requires addressing several critical technical challenges. We briefly highlight several here.

\subsection{Discovery and Interfaces}

Composing autonomous discovery processes requires first identifying
the agents that are necessary and perhaps evolving the set of participating agents over time. New discovery capabilities are needed 
to allow agents to discover other agents, determine what those agents can do, and how well they can do it.  Agents may also consider other aspects such as cost, safety, reliability, ethics, and so forth.   A second concern then is how agents may interact with one another, for example, via secure and self-descriptive interfaces. 
The significant prior work on actors provides a general basis for such interfaces; however, recent work with LLM and chat interfaces, agent-based frameworks like AutoGen, and remote computing, robot, and data interfaces must also be considered.

\subsection{Access Control and Sharing}
By definition, agents will interact with other agents. In many cases
these agents will span resources, facilities, institutions, regions, and even countries. As a result, it is critical that rules and regulations are followed and enforced, for example, limiting what resources may be used, how those resources are shared, and what resources can be used for what purposes. Overarching policies, defined by humans or machines, can be used to define these rules, and agents must incorporate mechanisms to enforce these policies. 

\subsection{Infrastructure}
The discovery process may include a broad range of resources---computation, data, models, people---that span locations and administrative domains. Making effective use of these resources
represents a considerable hurdle with respect to interfaces, policies, 
and usage requirements. Further, many scientific problems involve diverse data types (e.g., images, text, numerical data, sensor readings), unique instruments (e.g., microscopes, telescopes, gene sequencers), hardware (clusters, edge sensors, AI accelerators), and model types that must be leveraged by agents.  A final concern is agent resilience as agents may fail in myriad ways. Detecting, abstracting, and recovering from failures is a long standing challenge in distributed computing which is amplified by deployment of autonomous agents. 

\subsection{Agent Mobility}
Multi-agent systems for autonomous discovery will rarely be static. Agents must be deployed, terminated, replaced, scaled, and on occasion, moved between locations. Such mobility must be a cornerstone of the infrastructure and communication methods described above. Prior work in mobile agents~\cite{pham98mobileagents}, often referred to as mobile code,  provides a foundation for research in this area, considering, for example, code portability, ad hoc networking, and communication patterns.

\subsection{Provenance and Reproducibility}
The scientific method is one of proof and validation. It is critical that others be able to not only understand what has been done, but also to validate the results and methods applied, and to reproduce and extend those methods.  The fact that learning agents may be prone to opaque decision making necessitates efforts focused on interpretability  and explainability of individual decisions. However, considering  the nascent state of scientific reproducibility, there is an opportunity for agentic discovery to rapidly improve the verifiability and reproducibility or research processes by integrating provenance as a first-class citizen in the discovery lifecycle. Such capabilities may rely on verifiable ledgers to document processes and decision making, the ability to introspect agents and their behaviors, and methods to reuse agents in different settings to reproduce results.

\section{Uncertainties and Risks}

The transition towards agent-driven discovery raises several uncertainties and risks.
The greatest uncertainty---as with any large-scale endeavor---is garnering buy-in from stakeholders, the
scientists, research institutions, and funding agencies that will devote time and resources towards enabling this future.
We discuss such risks that, if ignored, will heighten stakeholder uncertainty.

Prior periods of research into agents, as mentioned earlier, have seen mixed success, and it is not unreasonable to be weary of history repeating itself.
Research then, just as it is now, was ambitious and the reasons for past failure were complex: AI methods were primitive, software protocols were more fragmented, and hardware limitations hindered innovation.
We believe that we have reached an inflection point where research in AI, software, and hardware has advanced to a point where their integration in the form of agents is feasible and will present a value add to science.
Not least, lessons learned from past failures will inform future decisions.

Worse than absolute failure or success is the illusion of success.
Autonomous systems are susceptible to security and safety risks, including adversarial attacks that could manipulate results or physical safety when autonomous actions are taken in the real world.
These risks are not unique to agents but require deference as ambition and scale increase.
In a similar vein, bias in any AI system must be characterized and controlled for, otherwise results may unknowingly be invalid.
For example, as in our MOF design application, biases in training data could lead to skewed discoveries that favor certain materials or methods over others.

Scientific discovery thrives on collaboration, yet autonomous workflows may inadvertently promote isolation if proprietary models or siloed datasets dominate research.
Open-source initiatives, shared repositories, and cross-disciplinary partnerships must be encouraged to prevent fragmentation and maximize the collective impact of agents.
Collaborations between human researchers and autonomous agents will pose new challenges to determining proper attribution of outcomes (just as we have seen with LLMs and writing).
Traditional academic crediting mechanisms may need to evolve to fairly recognize the tightly coupled contributions of humans and AI.

\vspace{1ex}

\chapteri{T}he future of scientific discovery will involve collections of autonomous agents working collaboratively in discovery workflows that drive the scientific process. 
In the near term, we expect that individual components will be iteratively augmented and replaced by agents, reducing bottlenecks and accelerating discovery.
These incremental improvements will compound, leading to fully autonomous discovery processes within the next five to ten years. 
Although progress toward this vision has accelerated in recent years, and nascent success stories are emerging, the coming decade will bring a profound transformation in how scientific discovery unfolds.

\bibliographystyle{IEEEtran}
\bibliography{main.bib}

\begin{IEEEbiography}{J. Gregory Pauloski}{\,} is a PhD Candidate in the Department of Computer Science at the University of Chicago. His research focuses on building novel frameworks and programming models for high-performance computing, deep learning training, and AI for science. Pauloski received his Masters in Computer Science from the University of Chicago. He is a student member of the IEEE. Contact him at \url{jgpauloski@uchicago.edu}.
\end{IEEEbiography}

\begin{IEEEbiography}{Kyle Chard}{\,} is a Research Associate Professor in the Department of Computer Science at the University of Chicago and Researcher at Argonne National Laboratory. His research focus on systems problems, often in high performance parallel and distributed computing, with the aim to develop new techniques for scalable and efficient management and analysis of large data. Chard received his Ph.D. in Computer Science from Victoria University of Wellington. He is a member of the IEEE. Contact him at \url{chard@uchicago.edu}.
\end{IEEEbiography}

\begin{IEEEbiography}{Ian Foster}{\,} is a Senior Scientist, Distinguished Fellow, and Director of the Data Science and Learning Division at Argonne National Laboratory and the Arthur Holly Compton Distinguished Service Professor of Computer Science at the University of Chicago. His research deals with distributed, parallel, and data-intensive computing technologies, and innovative applications of those technologies to scientific problems in such domains as materials science, climate change, and biomedicine.
Foster received his Ph.D. in Computer Science from Imperial College London. He is a fellow of the AAAS, ACM, BCS, and IEEE and an Office of Science Distinguished Scientists Fellow. Contact him at \url{foster@uchicago.edu}.
\end{IEEEbiography}

\end{document}